\documentclass[
aps,
 ,twocolumn
]{revtex4}
\newcommand{\ket}[1]{|{#1}\rangle}
\newcommand{\bra}[1]{\langle{#1}|}
\usepackage{graphicx}
\usepackage{epsfig}
\usepackage{amsfonts}
\usepackage{amsmath}

\newcommand{\beq}{\begin{equation}}
\newcommand{\eeq}{\end{equation}}
\newcommand{\bqa}{\begin{eqnarray}}
\newcommand{\eqa}{\end{eqnarray}}
\newcommand{\nn}{\nonumber}

\newcommand{\erf}[1]{Eq.~(\ref{#1})}

\newcommand{\ip}[2]{\langle{#1}|{#2}\rangle}

\newcommand{\cu}[1]{\left\{ {#1} \right\}}
\newcommand{\ro}[1]{\left( {#1} \right)}
\newcommand{\an}[1]{\left\langle{#1}\right\rangle}
\newcommand{\st}[1]{\left|{#1}\right|}

\newcommand{\bbb}[1]{\hspace{-5ex}{\phantom{\an{X_{w}}}}_{#1}\!}

\begin{document}

\title {Quantum Non-demolition Measurements on Qubits}

\author
{T.C.Ralph$^{1,2}$, S.D.Bartlett$^{2}$,  J.L.O'Brien$^{1,2}$, G.J.Pryde$^{1,2}$ and H.M.Wiseman$^3$}

\affiliation {$~^1$Centre for Quantum Computer Technology,\\
$~^2$Department of Physics, University of Queensland, Brisbane, QLD 4072,
Australia\\
$~^3$Centre for Quantum Computer Technology,
School of Science, Griffith University, Brisbane 4111 Australia\\
email: ralph@physics.uq.edu.au}
%

\begin{abstract}

We discuss the characterization and properties of quantum non-demolition (QND) measurements on qubit systems. We introduce figures of merit which can be applied to systems of any Hilbert space dimension thus providing universal criteria for characterizing QND measurements. We discuss the controlled-NOT gate and an optical implementation as examples of QND devices for qubits. We also discuss the QND measurement of weak values.

\end{abstract}

\maketitle

\section{Introduction}

The act of measuring a quantum system to acquire information about it
must necessarily disturb the system.  Quantum non-demolition (QND)
measurements~\cite{Gra98} allow for the measurement of an observable
of a quantum system without introducing a back-action on this
observable due to the measurement itself.  QND measurements explore
the fundamental limitations of measurement and may prove useful in
gravity wave detection \cite{Bra80}, telecommunications \cite{Lev93} and quantum control \cite{Wis95}.

The traditional domain of experimental QND measurements is
continuous-variable (CV) quantum optics \cite{Roc97,Buc02}. 
CV QND measurements are performed
using only Gaussian states (those states of the electro-magnetic field with a Gaussian Wigner
function), working with quadrature components of the field
proportional to number and phase in a linearised regime. They have
been characterized by considering the signal to noise transfer and
conditional variances between various combinations of the input,
output and measurement output of the device. These are known as
``T-V" measures~\cite{Poi94}.

In contrast, discrete variable quantum optics typically deals with two level quantum systems such as the polarization states of single photons. Quantum bits or ``qubits" can be carried by such systems. Progress in the field of quantum information, in particular in the
realization of two qubit gates, has opened a new domain in which QND
measurements can be demonstrated. Indeed QND measurements are
critical to many key quantum information protocols, such as error correction \cite{Ste96}, and enable
new computation models \cite{Dur03}. However, in
this new domain of low-dimensional quantum systems with arbitrary
superposition states, the standard measures used to characterize CV
QND measurements are no longer applicable.

Until recently it was only in the domain of cavity quantum electrodynamics that interactions sufficiently strong as to probe the qubit domain could be achieved with optical fields \cite{Nog99, Tur95}. However
the work of Knill, Laflamme and Milburn~\cite{KLM01} introduced the technique of measurement induced non-linearities and
led to proposals for non-deterministic realizations of QND measurements for traveling fields \cite{Kok02}. In these schemes the non-linearity is induced through photon counting measurements made on ancilla modes which have interacted with the system modes via linear optics.
A demonstration
of a QND measurement on a single photonic qubit was recently made by
Pryde {\it et al}~\cite{Pry04}. In this paper we investigate the character,
characterization, an optical implementation and a fundamental 
application of QND measurements on qubits.

We begin in the next section by describing the basic features that a QND
measurement should display. We then propose quantitative measures by
which the quality of any QND measurement can be assessed. We consider qubit systems primarily but also
discuss the application of these measures to systems of any
dimension. In section \ref{backaction} we consider the trade-off between the accuracy of the QND measurement and its inevitable back-action on the conjugate observable to that being measured. 
In section \ref{cnot} we discuss the example of the controlled-not
(CNOT) gate and show how it can be used to to make generalized QND
measurement of arbitrary strength. How such measurements can be
implemented in optics is described in section \ref{optics}. In
section \ref{weak} we discuss the domain of weak valued measurements and
propose experiments which would highlight some fundamental
peculiarities of quantum mechanics. 

\section{Fidelity measures for QND measurements}
\label{section2}

\begin{figure}
\begin{center}
\includegraphics[width=5.0cm]{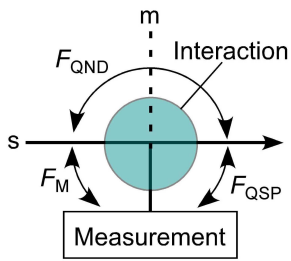}
\caption{A schematic of a QND measurement. After interaction with the signal, {\it s}, the meter, {\it m}, carries information about the signal which can be read out by a measurement. The performance against the requirements (1.-3.) can be assessed by measuring the indicated correlations.}
\label{qnd}
\end{center}
\end{figure}

A measurement device takes a quantum system in an input state, described in
general by the density matrix $\hat{\rho}$, and via an interaction
yields a classical measurement outcome, $i$, of some particular observable. The quantum system is left in the corresponding
output state $\hat{\rho}'_i$.  To be considered a QND measurement,
the device should satisfy three inter-related physical requirements \cite{Gra98}:
\begin{enumerate}
\item \emph{The measurement result should be correctly correlated with
     the state of the input.}  For example, if the input state is an
   eigenstate of the observable being measured, then in an ideal QND
   measurement the measurement outcome corresponding to this eigenstate should occur with
   certainty.
\item \emph{The measurement should not alter the observable being
     measured.}  For example, an eigenstate should be left in the same
   eigenstate by the measurement.
\item \emph{Repeated measurements should give the same result.}  In
   other words, the QND measurement should be a good quantum state
   preparation (QSP) device, and should output the eigenstate
   corresponding to the measurement result.
\end{enumerate}
In a realistic QND measurement scheme, these requirements will not be perfectly satisfied.  In the following we propose quality
measures for each of the above requirements that apply to arbitrary
distributions of input states. A schematic representation of a QND measurement is shown in Fig.\ref{qnd}.

\subsection{Quantifying performance with fidelities}

A QND measurement can be tested relative to the criteria (1.-3.) by
performing repeated measurements of a set of known signal input states
$\{\hat{\rho}\}$.  Let $\{|\psi_i\rangle, i=1,\ldots,d\}$ be a basis
of eigenstates of the measurement of a qudit system with dimension
$d$.  There are three relevant probability distributions: $p^{\rm in}$
of the signal input, which consist of the diagonal elements of the
signal input density matrix $p^{\rm in}_i = \langle \psi_i
|\hat{\rho}|\psi_i \rangle$ in the basis of eigenstates of the
measurement; the distribution $p^{\rm out}$ of the signal output,
which consists of the diagonal elements of the signal output density
matrix $p^{\rm out}_i = \langle \psi_i |\hat{\rho}'|\psi_i \rangle$;
and the distribution $p^{\rm m}$ of the measurement statistics of the
meter.  These distributions are all functions of the signal input state.
The requirements (1.-3.) demand \emph{correlations} between these
distributions.

To quantify the performance of a QND measurement, we define measures
that can be applied to all input states. These measures each compare
two probability distributions $p$ and $q$ over the measurement
outcomes $i$, using the (classical) fidelity
\begin{equation}
   F(p,q)=\Bigl(\sum_i\sqrt{p_i q_i} \Bigr)^2.
\end{equation}
Note that $F=1$ for identical distributions, whereas $F = d^{-1}$ for
uncorrelated distributions (for example, comparing a lowest entropy
distribution $\{1,0,0,\ldots,0 \}$ and the highest entropy
distribution $\{d^{-1},d^{-1},\ldots,d^{-1}\}$).  For the special case
of $d=2$, we also have that $F = 0$ for anti-correlated distributions
(for example, $\{1,0\}$ and $\{0,1\}$).  As with the probability
distributions, these fidelities are also functions of the signal input
state. 

\subsection*{Requirement 1: Measurement fidelity}

The first requirement demands that the measurement result is
correlated with the state of the input.  A device can be tested
against this requirement by measuring a set of known states
$\{\hat{\rho}\}$ and analyzing the resulting statistics.  For example,
consider tests involving signal input states that are eigenstates of
the observable being measured.  Comparing the input distribution
$p^{\rm in}$ (which, in this case, are lowest entropy distributions of
the form $\{0,0,\ldots,1,\ldots,0\}$) with the measurement
distribution $p^{\rm m}$ (consisting of the probabilities of measuring
the result $i$) quantifies the correlation between the input and the
measurement result.  However, we note that a QND measurement device is
also expected to reproduce the expected measurement statistics for any
state, including those that are superpositions of eigenstates, and
thus it is necessary to analyze the performance for such
non-eigenstate inputs as well.

To quantify the performance of a QND measurement against requirement
1., we define the \emph{measurement fidelity} for the input state
$\hat{\rho}$ to be
\begin{equation}
   \label{eq:MeasurementFidelity}
   F_{\rm M} = F(p^{\rm in},p^{\rm m}) \, ,
\end{equation}
which gives the overlap between the signal input and measurement
distributions.

As an illustrative example, consider a device where the measurements
results are uniformly random and completely uncorrelated with the
input states.  The measurement statistics are then $p^{\rm m} =
\{d^{-1},d^{-1},\ldots,d^{-1}\}$.  The resulting measurement fidelity
will then range from $d^{-1} \leq F_{\rm M} \leq 1$, where the lower
bound is obtained for eigenstate inputs and the upper bound is
obtained for maximally mixed states or equally-weighted superpositions
of all eigenstates.  Note that, for qubits, a measurement fidelity of
$0$ is only obtained if the measurement results are completely
anti-correlated with the input state.

The measurement fidelity $F_{\rm M}$ can be used to quantify the
performance as a measurement device for any signal input state.  Of particular interest are the measurement fidelities for the
eigenstates, $|\psi_i\rangle$, of the observable, which should ideally
give $F_M = 1$.  It is also important to ensure that all other
superposition states produce the correct statistics, and thus the
measurement fidelity should be measured for a representative set of
states, ideally one which spans the system Hilbert space.

\subsection*{Requirement 2:  QND fidelity}

The second requirement is that the measurement does not disturb the
observable to be measured, i.e., that the measurement is QND.  Note,
however, that if the input state is not an eigenstate of the
observable being measured, the state must necessarily be altered by
the measurement process since the measurement should, ideally, project the signal into an eigenstate.  Therefore the signal output of an ideal QND measurement device (when we trace out the meter state) will be in a mixed state where the
coherences, $\rho_{ij} = \langle \psi_i|\hat{\rho}|\psi_j\rangle$,
$i\neq j$, between different eigenstates have been removed, whilst leaving the diagonal
elements unaltered.

Thus the distribution $p^{\rm in}$ should be preserved by the measurement,
i.e., it should be identical to the distribution $p^{\rm out}$.  We
compare these two distributions, again using the classical fidelity as
a measure, and define the \emph{QND fidelity} for the input state
$\hat{\rho}$ to be
\begin{equation}
   \label{eq:QNDFidelity}
   F_{\rm QND} = F(p^{\rm in}, p^{\rm out}) \, .
\end{equation}
This measure ranges from $0$ to $1$, yielding $1$ only if the
distributions are identical.  If, for example, the measurement always
produces an eigenstate output but alters (``flips'' in the $d=2$
qubit case) the value of $i$ each time, the resulting QND fidelity
would be zero.

For input states that are eigenstates of the measurement, the QND
fidelity characterizes how well the measurement preserves the measured
observable.  For input states that are superpositions of eigenstates,
the QND fidelity characterizes how well the average populations are preserved.

\subsection*{Requirement 3:  QSP fidelity}

Finally, for a QND measurement we also require that the output state
should be the eigenstate $|\psi_i\rangle$ after obtaining the measurement
result $i$.  Thus, a measure of quality is needed to characterize how
well the output state compares to $|\psi_i\rangle$, i.e., how well the
measurement acts as a QSP device.

Let $p_{\ket{i}|i}^{\rm out} $ denote the \emph{conditional} probability of finding
the signal output state to be $\ket{i}$ given that the QND measurement gave
the measurement result $i$.  We define the \emph{QSP fidelity}
\begin{equation}
   F_{\rm QSP}=\sum_i p_i^{\rm m} p_{\ket{i}|i}^{\rm out} \, ,
\end{equation}
which is an average fidelity (averaged over all possible measurement
outcomes) between the expected and observed conditional probability
distributions.  This QSP fidelity has the desirable property that it
ranges from $0$ (if the output state is always orthogonal to the
desired outcome) to $1$ (if they are equal).  If the outcome is
uncorrelated with $i$, the QSP fidelity would have a value of
$d^{-1}$.

For qubits, the QSP fidelity is also known as the \emph{likelihood},
$L$, \cite{wo-prd-19-473} of measuring the signal to be $i$ given the
meter outcome $i$. We will use this interpretation in the next section to quantify the back action of a QND measurement.

The most relevant QSP fidelity is obtained when the input is
completely unknown, i.e., a completely mixed state.
Then this QSP fidelity characterizes how well the measurement device
prepares a definite quantum state (labeled by $i$) given no prior
knowledge.

\subsection*{Relation between quality measures}

As noted above, these three fidelities $F_{\rm M}$, $F_{\rm QND}$ and
$F_{\rm QSP}$ are interrelated and not independent.  Each
is used to compare two of the three probability distributions relevant
to a QND measurement for a given input state: the distribution $p^{\rm
   in}$ of input probabilities, the distribution $p^{\rm out}$ of
output probabilities, and the distribution $p^{\rm m}$ of measurement
outcomes.  Any two of these fidelity measures thus sets bounds on the
third.

\subsection{Comparing with CV measures}

Traditionally, QND measurements have been realised experimentally in
quantum optics using Gaussian states (coherent and squeezed states
with Gaussian Wigner functions) with a sufficiently large average
photon number to allow for a linearized treatment of the quantum
noise \cite{note1}. Standard quality measures for this particular
type of QND measurement
have been proposed~\cite{Poi94}.  In the following, we show how these
CV quality measures compare with the general fidelities proposed above
(which apply to any $d$-dimensional system). We note that the available input states in the CV domain do not allow for a complete investigation of the fidelities described above. For example, it is not possible to  inject eigenstates of the QND observable because quadrature eigenstates are unphysical.

The CV quality measures are defined in terms of correlations between
the input and output states of the quantum system being measured (the
signal) and the input and output states of the quantum system used as
a meter.  Consider a QND observable $\hat{O}$. An example for a qubit system would be the Stokes operator $\hat{S}_1 = \ket{H}\bra{H} - \ket{V}\bra{V} $ and for a CV system the quadrature amplitude fluctuation $\delta \hat X = \hat X - \langle
X\rangle$, where $\hat X=\hat a + \hat a^{\dagger}$ and $\hat a$ is the normal boson annihilation operator. Correlation functions are defined as follows.  For two systems $A$ and $B$ with QND observable $\hat
{O}_A$ and $\hat{O}_B$ respectively, the correlation function between them is
defined as
\begin{equation}
    C^2(O_{A}O_{B}) = \frac{\tfrac{1}{4}|\langle \hat {O}_{A} \hat {O}_{B}
      \rangle + \langle \hat {O}_{B} \hat {O}_{A}
      \rangle|^2}{\langle \hat {O}_{A}^{2} \rangle \langle \hat
      {O}_{B}^{2} \rangle} \, .
\end{equation}
Thus, in CV systems, the correlation function between the quadratures of two systems
$\delta X_A$, $\delta X_B$ is
\begin{equation}
   \label{eq:CorrelationCoefficient}
   C^2(\delta X_A\,\delta X_B) = \frac{\tfrac{1}{4}|\langle \delta X_A \delta
   X_B\rangle + \langle \delta X_B \delta X_A\rangle|^2}{\langle \delta
   X_A^2\rangle\langle \delta X_B^2  \rangle} \, ,
\end{equation}
which provides a measure of the correlation of the two fluctuations,
ranging from $0$ (uncorrelated) to $1$ (perfectly correlated).  The
quality measures for a CV QND measurement scheme against the
criteria listed earlier in this section are:

\textbf{Quality of measurement.}  The correlation between the input
state of the system with fluctuations $\delta X_{\rm s}^{\rm in}$ and
the output state of the meter with fluctuations $\delta M^{\rm out}$
is given by the correlation coefficient
\begin{equation}
   \label{eq:MeasurementCorrelation}
   C^2_{\rm m} = C^2(\delta X_{\rm s}^{\rm in}\,\delta M^{\rm out})\, .
\end{equation}
This quantity ranges from $0$ (for no correlation between the system
and the measurement) to $1$ (for a perfect measurement); it is thus
quantitatively comparable to the measurement fidelity $F_{\rm M}$ when
applied to the infinite-dimensional CV system, which then also equals
zero for no correlation and unity for perfect correlation. In practice, the correlation $C^2_{\rm m}$ is not easily measurable in an experiment, and thus the signal to noise
transfer coefficient, $T_{\rm M}$, is typically used \cite{Gra98,Poi94}. $T_M$ is the ratio of the signal to noise of the meter output to the signal to noise of the signal input. This
transfer coefficient can be related to $C^2_{\rm m}$ if one imposes restrictions on the input states; i.e. only Gaussian states and particular choices of squeezing axes (major and minor axes of the ellipse). Given these restrictions there is a direct relationship between $F_{\rm M}$ and $T_{\rm M}$ given by
\begin{equation}
F_{\rm M} = \sqrt{{{2  T_{\rm M}}\over{1+T_{\rm M}}}} .
\end{equation}

\textbf{Quality of QND.} The correlation between the input and output
states of the system is given by the correlation coefficient
\begin{equation}
   \label{eq:QNDCorrelation}
   C^2_{\rm s} = C^2(\delta X_{\rm s}^{\rm in}\,\delta X_{\rm s}^{\rm
   out})\, .
\end{equation}
This quantity ranges from $0$ to $1$, and $C^2_{\rm s}=1$ if the
observable $X$ is not disturbed or degraded.  The QND fidelity applied
to this CV system is quantitatively comparable to $C^2_{\rm s}$. As for $C^2_{\rm m}$, $C^2_{\rm s}$ is not easily measurable in practice so the ratio of the signal output signal to noise to that of the input, the signal transfer coefficient $T_{\rm S}$, is used. Given the restrictions outlined above there is a direct relationship between $F_{\rm QND}$ and $T_{\rm S}$ given by
\begin{equation}
F_{\rm QND} = \sqrt{{{2  T_{\rm S}}\over{1+T_{\rm S}}}} .
\end{equation} 

\textbf{Quality of QSP.}  The conditional variance $V_{\rm
   s|m}$, defined in terms of the fluctuations $\delta X_{\rm s}^{\rm
   out}$ and $\delta M^{\rm out}$ of the output state of the system and
the output state of the meter, respectively, is
\begin{equation}
   \label{eq:ConditionalVariance}
   V_{\rm s|m} =
   \langle \delta
   {X_{\rm s}^{\rm out}}^2 \rangle (1 - C^2 (\delta X_{\rm s}^{\rm
   out}\,\delta M^{\rm out})) \, .
\end{equation}
This quantity is defined such that $V_{\rm s|m}=0$ corresponds to
perfect correlation between the system and the meter, and $V_{\rm
   s|m}<1$ indicates conditional squeezing (quantum correlations).  Due
to the inclusion of $\langle \delta {X_{\rm s}^{\rm out}}^2 \rangle$
in $V_{\rm s|m}$ as a reference to the shot-noise level, it is not
possible to directly compare this quantity to the QSP fidelity.  However, if we
tried to go the other way and define the QSP fidelity in the CV
regime, we find that $\overline{F}_{\rm QSP}=0$ due to the continuous
spectrum of the measurement outcome. To compare directly with CV
experiments, the QSP performance for both finite and CV systems could
also be quantified by the correlation function between the signal and
meter. To this end we note that for qubits $C^2(O_{m}O_{s}) = 2 {F}_{\rm QSP} -1$.

\section{Back-action on the Conjugate Variable}
\label{backaction}

So far we have been assuming that the aim of the QND measurement is to make a non-destructive {\it projective} measurement of the system. However such measurements are not the only possibility. We may wish to make a generalized measurement \cite{Nie00} of our system; one that extracts only partial information about the observable in question. In such a situation an additional figure of merit arises: the extent to which the measurement decoheres the conjugate observable to that being measured. For a projective measurement, we expect the conjugate observable to be completely decohered. However, for a partial measurement, the back-action on the conjugate variable need only be sufficient to satisfy complementarity. We can quantify this trade-off for qubits by comparing the distinguishability of the eigenstates of the QND observable via measurements at the meter output with the distinguishability of the eigenstates of the conjugate to the QND observable via measurements at the signal output. 

Recall first that the QSP fidelity describes the likelihood, $L$, that the meter outcome coincides with the signal outcome. We can thus define the distinguishability of the eigenstates of the QND observable, based on meter measurements as: $K = 2 L - 1$. Now suppose eigenstates of a conjugate observable are injected at the signal input and measurements in the same conjugate basis are made at the signal output. We can define the distinguishability of the conjugate eigenstates based on signal measurements as $\bar K = 2 P_{c} - 1$ where $P_{c}$ is the probability that the signal out detector correctly identifies the injected eigenstate. Complementarity would suggest that the mutual distinguishability of the two observables should be bounded. This was quantified by Englert \cite{eng96} who showed that
\begin{equation}
K^2 + \bar K^2 \le 1.
\label{ba}
\end{equation}
For an ideal projective QND measurement we would expect $K=1$ and so $\bar K = 0$. For an ideal generalized QND measurement we might have $K < 1$ but would expect the inequality of Eq.\ref{ba} to be saturated. We will refer to a QND measurement in which a reduction in $K$ does not result in a corresponding increase in $\bar K$ as an {\it incoherent} QND measurement. An example of an incoherent QND measurement is a ``measure and re-create" procedure in which a destructive measurement is made of the system and then a new signal state is generated based on the result of the measurement. Most quantum information applications require coherent QND measurements.

For CV systems back-action on the conjugate variable can be quantified by the generalized uncertainty principle \cite{Art88} which requires that 
\begin{equation}
V_{\rm s|m} V_{conj} \ge 1
\end{equation}
where $V_{conj}$ is the variance found in a direct measurement of the conjugate variable on the signal output.

\section{Performing QND measurements of qubits with a CNOT}
\label{cnot}

We now discuss performing QND measurements in an arbitrary basis (i.e. of an arbitrary observable) and
of arbitrary strength on qubits
using a controlled-NOT (CNOT) gate. A strong (projective)
QND measurement of a qubit in the computational basis
can be made with a CNOT in the following
way:
\begin{enumerate}
\item the target qubit plays the role of the meter, it is prepared
in the logical zero state, $\ket{\bf 0}$;
\item the signal, in an arbitrary state $\alpha \ket{\bf 0} + \beta
\ket{\bf 1}$, is the control qubit;
\item the gate is run producing the output state:
\begin{equation}
\alpha
\ket{\bf 0}_{s}\ket{\bf 0}_{m} + \beta
\ket{\bf 1}_{s}\ket{\bf 1}_{m},
\label{gate1}
\end{equation}
where $s$ labels the signal (ie the control)
output and $m$ labels the meter (ie the target) output;
\item the meter is measured in the computational basis.
\end{enumerate}
With probability $|\alpha|^{2}$ the meter measurement will give the
result ``0" and the signal output state will be left in the state
$\ket{\bf 0}$. Similarly, with probability $|\beta|^{2}$ the meter
measurement will give the result ``1" and the signal state will be
left in the state $\ket{\bf 1}$.

The reduced state of both the signal and the meter outputs, if we trace over the other, is
\begin{equation}
     \rho_{m/s} = |\alpha|^{2} \ket{\bf 0}\bra{\bf 0} + |\beta|^{2} \ket{\bf 1}\bra{\bf 1}.
\end {equation}
Thus the probability distributions for measurements in the computational basis
will be identical for the input and outputs:
\begin{equation}
p^{in} = p^{out} = p^{m} = \{ |\alpha|^{2}, |\beta|^{2} \}
\end {equation}
and so against our first two criteria we obtain
$F_{\rm M} = F_{\rm QND} = 1$, for all input states. The conditional
probability that a ``0'' (``1'') result at the meter results in a
``0'' (``1'') result at the signal out is seen from Eq.\ref{gate1} to
be unity. Thus we also have $F_{\rm QSP} = 1$. The CNOT then gives an ideal QND measurement.

Similarly the correlation between the signal and meter outputs is
\begin{eqnarray}
    C^2(Z_{s}Z_{m}) & = & \frac{|\langle \hat {Z}_{s} \hat {Z}_{m}
     \rangle|^2}{\langle \hat {Z}_{s}^{2} \rangle \langle \hat
      {Z}_{m}^{2} \rangle} \nonumber \\
      & = & 1
\end{eqnarray}
where $\hat {Z}_{i} = \ket{\bf 0}\bra{\bf 0}_{i} - \ket{\bf 1}\bra{\bf 1}_{i}$ is a
computational basis measurement on the ith output and the expectation
value is taken over the output state (Eq.\ref{gate1}).

The QND measurement basis can be altered simply by rotating the
signal qubit from which ever basis the measurement is desired for,
into the computational basis, applying the above protocol, then
rotating back to the original basis after. 

We now examine altering the strength of the QND measurement . Consider first what occurs if the protocol is followed as
above except that we prepare the meter in the diagonal state $(\ket{\bf 0} +
\ket{\bf 1})/\sqrt{2}$. The output state of the gate is then
\begin{equation}
     (\alpha \ket{\bf 0} + \beta
\ket{\bf 1})_{s} {{1}\over{\sqrt{2}}}(\ket{\bf 0} +
\ket{\bf 1})_{m}.
\end{equation}
There is now no correlation between the meter and the
signal outputs; tracing over the meter leaves the signal in its
original state, the QND measurement has been``turned off". The
probability distributions in the computational basis are now
\begin{eqnarray}
p^{in} & = & p^{out} = \{ |\alpha|^{2}, |\beta|^{2} \}\nonumber\\
p^{m} & = & \Big\{{{1}\over{2}},{{1}\over{2}}\Big\}
\end {eqnarray}
and we
find $F_{\rm M} = 1/2 +
\sqrt{|\alpha|^{2}|\beta|^{2}}$. For eigenstate inputs we calculate $F_{\rm M} = 1/2$,
indicating no correlation. On the other-hand the signal state has not
been disturbed and thus $F_{\rm QND} = 1$. The conditional probabilities are
\begin{equation}
p_{0,1}^{out}|_{0,1} =  \Big\{ {{1}\over{2}}|\alpha|^{2},
{{1}\over{2}}|\beta|^{2} \Big\}
\end {equation}
giving $F_{\rm QSP} = 1/2$ and also $C^2(Z_{s}Z_{m}) =
0$, both indicating no correlation between the meter and signal output.

More interestingly, let us now see what occurs if the meter is
prepared in a state lying in between that producing a strong QND
result and that producing no measurement. We thus prepare the meter in the state
$\gamma \ket{\bf 0} + \bar \gamma \ket{\bf 1}$ where $\gamma$ is a
real number between $1/\sqrt{2}$ and $1$ and $\bar \gamma = \sqrt{1- \gamma^{2}} $. The joint meter signal output state is
\begin{eqnarray}
\alpha \gamma
\ket{\bf 0}_{s}\ket{\bf 0}_{m} & + & \alpha \bar \gamma
\ket{\bf 0}_{s}\ket{\bf 1}_{m} \nonumber\\
& + &  \beta \gamma
\ket{\bf 1}_{s}\ket{\bf 1}_{m} + \beta \bar \gamma
\ket{\bf 1}_{s}\ket{\bf 0}_{m}.
\label{gate2}
\end{eqnarray}
The reduced density operators for the signal and meter outputs are:
\begin{eqnarray}
\rho_{s} & = & (\alpha \gamma
\ket{\bf 0} + \beta \bar \gamma \ket{\bf 1}) (\alpha^\ast \gamma
\bra{\bf 0} + \beta^\ast \bar \gamma \bra{\bf 1})  \nonumber\\
& + & (\alpha \bar \gamma
\ket{\bf 0} + \beta \gamma
\ket{\bf 1}) (\alpha^\ast \bar \gamma
\bra{\bf 0} + \beta^\ast \gamma
\bra{\bf 1}) \nonumber\\
\rho_{m} & = &  (\alpha \gamma
\ket{\bf 0} + \alpha \bar \gamma \ket{\bf 1}) (\alpha^\ast \gamma
\bra{\bf 0} + \alpha^\ast \bar \gamma \bra{\bf 1}) \nonumber\\
& + & (\beta \bar \gamma
\ket{\bf 0} + \beta \gamma
\ket{\bf 1})  (\beta^\ast \bar \gamma
\bra{\bf 0} + \beta^\ast \gamma
\bra{\bf 1}) 
\label{gqnd}
\end {eqnarray}
The relevant
probability distributions are now
\begin{eqnarray}
p^{in} & = & p^{out} = \{ |\alpha|^{2}, |\beta|^{2} \}\nonumber\\
p^{m} & = & \{|\alpha|^{2} \gamma^{2} + |\beta|^{2}\bar \gamma^2, \nonumber\\
& & \;\;\; |\beta|^{2} \gamma^{2} + |\alpha|^{2}\bar \gamma^2\}.
\end {eqnarray}
Now 
\begin{equation}
F_{\rm M} = (\sqrt{|\alpha|^{2}(|\alpha|^{2} \gamma^{2} +
|\beta|^{2}\bar \gamma^2)} + \sqrt{|\beta|^{2}(|\beta|^{2} \gamma^{2} +
|\alpha|^{2}\bar \gamma^2)}\;)^{2}.
\end{equation}
Eigenstate inputs yield $F_{\rm
M} = \gamma^{2}$, thus we can smoothly vary between an ideal QND
measurement ($\gamma = 1$, $F_{\rm M} = 1$) and no measurment
($\gamma = 1/\sqrt{2}$, $F_{\rm M} = 1/2$) by changing $\gamma$.
The measurement maintains $F_{\rm QND} = 1$ for all $\gamma$. The conditional
probabilities are
\begin{equation}
p_{0,1}^{out}|_{0,1} =  \{|\alpha|^{2} \gamma^{2},
|\beta|^{2} \gamma^{2} \}
\end {equation}
and so $F_{\rm QSP} = \gamma^{2}$ and also $C^2(Z_{s}Z_{m}) =
2 \gamma^{2} - 1$. The correlation between the outputs can be smoothly
varied between correlated and uncorrelated by tuning the value of
$\gamma$.

Running in this mode of operation, the CNOT gate is performing a generalized measurement. To see if this is coherent we need to evaluate Eq.\ref{ba}. We have $K = 2 \gamma^{2} - 1$. Setting $\alpha = \beta = 1/\sqrt{2}$ (a diagonal signal input state) and asking for the probability, $P_{c}$, that $\rho_{s}$ is measured in a diagonal output state gives the result (from Eq.\ref{gqnd}) $P_{c} = \gamma \bar \gamma + 1/2$. Then $\bar K = 2 \gamma \bar \gamma$ and $K^2 + \bar K^2 = 1$ as required for coherent operation.

\section{A Non-deterministic Optical Implementation}
\label{optics}

We now consider an optical implementation of a qubit QND measurement which is based on linear optical interactions plus a measurement induced non-linearity. Such an implementation is non-deterministic and, with current technological limitations, relies on coincidence detection. 
An experimental demonstration of this scheme was presented in Ref.~\cite{Pry04}. Here we briefly review the scheme then discuss its characterization in more detail .

\subsection{QND measurement of photon polarisation}

\begin{figure}
\begin{center}
\includegraphics[width=7.5cm]{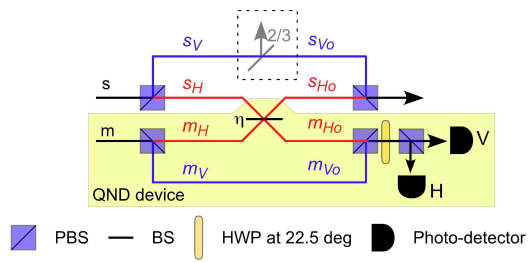}
\caption{A schematic circuit for the realisation of a QND measurement of the polarisation of a single photon. Input signal, $s$, and meter, $m$, modes containing single photon states are split into different spatial rails by polarizing beamsplitters. The horizontal signal and meter modes are mixed on a beamsplitter leading to non-classical interference. Detection of a single meter photon (and absence of a photon at the signal dump port) heralds success of the QND measurement. See text for details.}
\label{schematic}
\end{center}
\end{figure}

The aim of our optical QND device is to imprint information about the polarisation of a single signal photon onto the polarisation state of a single meter photon. The polarisation degree of freedom of a single photon is a 2-dimensional system: a qubit. The net effect of this device is to perform a projective (or generalized) measurement of polarisation on a single photon which is then free propagating after this measurement. We describe a scheme for realising such a measurement non-deterministically using linear optics and single photon measurement.
The optical circuit is shown schematically in Fig.~\ref{schematic}. The scheme works on similar principles to those outlined in section \ref{cnot} as its implementation is closely related to the non-deterministic CNOT gate described in  \cite{ral02,hof02,job03}.

In the circuit of Fig.~\ref{schematic} a signal and a meter photon are each input from the left. The polarisation modes are separated into spatial modes using a polarising beam splitter (PBS). In the language of optical quantum information the qubit stored in each photon is converted from a polarisation encoding into a spatial encoding. The $s_H$ and $m_H$ modes then interfere non-classically on a beam splitter (BS) with reflectivity $\eta$ ($\eta$BS). Non-classical interference between single photons on a BS arises from interference of indistinguishable amplitudes \cite{HOM}. The four spatial modes are then recombined at a second pair of PBSs as indicated.

When two photons are input into a $\frac{1}{2}$BS the probability of detecting a single photon at each of the outputs is given by the absolute square of the sum of the indistinguishable amplitudes. The two indistinguishable amplitudes correspond to cases where both photons are reflected or both photons are transmitted. Because there is a total $\pi$ phase shift on reflection, these amplitudes cancel one another, and the probability to detect a single photon at each output is zero. This is in contrast to our classical expectation which would lead us to predict that the probability of such an event is $\frac{1}{2}$. For an arbitrary $\eta$BS this same effect means that the probability of detecting a single photon in each output port is simultaneously reduced by the amount:
\begin{equation}
\label{ }
\frac{(1-2\eta)^2}{(1-\eta)^2+\eta^2}.
\end{equation}
In the following discussion we consider a BS where the phase shift on reflection is $\pi$ on one side and zero on the other without loss of generality.

If we consider just the part of the circuit in Fig.~\ref{schematic} between the two pairs of PBSs, then the Heisenberg equations relating the input modes to the output modes of the circuit are:
\begin{eqnarray}
&& s_{V_O}=s_V \\
&& s_{H_O}=\sqrt{\eta}\; s_H+\sqrt{1-\eta}\; m_H\\
&& m_{H_O}=\sqrt{1-\eta}\; s_H-\sqrt{\eta}\; m_H\\
&& m_{V_O}=m_V.
\end{eqnarray}
The goal of the device is that the meter photon interacts with the signal photon in such a way that subsequent measurement of the meter polarisation tells us the polarisation of the signal. We first consider what the output state of the circuit is for eigenstate signal inputs: $|H\rangle_s$ and $|V\rangle_s$. 

We compensate for loss in the meter $H$ component (due to the beam splitter) by adjusting the input  state of the meter according to
\begin{equation}
|D(\eta)\rangle_m\equiv\sqrt{\frac{1}{1 + \eta}}|H\rangle_m + \sqrt\frac{\eta}{1 + \eta}|V\rangle_m.
\label{d}
\end{equation}
This ensures that the meter exits with orthogonal polarizations for different eigenstates. In the case where there is no $s_H$ signal photon, the meter photon exits the meter port with probability $\sqrt{2\eta/(1 + \eta)}$. When this happens, the $H$ and $V$ components of the meter state are equal due to the $\sqrt{1-\eta}$ loss in the $m_H$ mode. The goal is then to introduce a $\pi$ phase shift in the $m_H$ mode conditional on there being a single photon in the $s_H$ mode so that the meter photon has equal $H$ and $V$ components with a $\pi$ phase shift and the meter outputs are orthogonal depending on whether there is a single photon in the $s_H$ mode.

For the signal input $|V\rangle_s$ the total input state (meter and signal) is
\begin{widetext}
\begin{equation}
|V\rangle_s|D(\eta)\rangle_m=|V\rangle_s\otimes\Big(\sqrt{\frac{1}{1+\eta}}|H\rangle_m+\sqrt{\frac{\eta}{1+\eta}}|H\rangle_m\Big)=\sqrt{\frac{1}{1+\eta}}|V\rangle_s|V\rangle_m+\sqrt{\frac{\eta}{1+\eta}}|V\rangle_s|H\rangle_m
\end{equation}
\end{widetext}
We can obtain the output directly using time reversal symmetry: 
\begin{equation}
|\phi^{V_s}_{out}\rangle=
\sqrt{\frac{\eta}{1+\eta}}|V\rangle_s(|V\rangle_m+|H\rangle_m)+\sqrt{\frac{1-\eta}{1+\eta}}|H\rangle_s|V\rangle_s.
\end{equation}
The first term corresponds to successful operation: one photon exits in each of the signal and meter modes. The second term corresponds to a failure: both photons leave in the signal mode. The probability of success for this signal input $P_V$ is dependent on $\eta$: 
\begin{equation}
P_V= \sqrt{\frac{2\eta}{1 + \eta}},
\end{equation}
and $P_V\rightarrow1$ as $\eta\rightarrow1$. When the circuit succeeds, the meter photon is diagonally polarised: $(|D\rangle\equiv|H\rangle+|V\rangle)/\sqrt{2}$ and the signal is vertically polarised. The half wave plate (HWP) in Fig. \ref{schematic} is oriented at 22.5$^{\circ}$ and rotates the meter from $|D\rangle$ to $|V\rangle$, so that the meter and signal then have the same polarisation. Measurement of the meter polarisation then gives the polarisation of the signal without destroying it. 

For the other signal eigenstate input $|H\rangle_s$, the total input state is
\begin{equation}
\label{}
|H\rangle_s|D(\eta)\rangle_m=\sqrt{\frac{1}{1+\eta}}|H\rangle_s|V\rangle_m+\sqrt{\frac{\eta}{1+\eta}}|H\rangle_s|H\rangle_m
\end{equation}
We again obtain the output directly using time reversal symmetry: 
\begin{equation}
|\phi^{H_s}_{out}\rangle=\sqrt{\frac{1}{1+\eta}}[(1-2\eta)|H\rangle_s|H\rangle_m-\eta|H\rangle_s|V\rangle_m]+...
\end{equation}
where the terms not shown are ones where two photons exit in either the signal or meter mode. The two coefficients are required to be equal (ie. $1-2\eta=\eta$) so that the meter output state is $(|H\rangle_m-|V\rangle_m)/\sqrt{2}\equiv|A\rangle$. This condition is only satisfied for $\eta=\frac{1}{3}$. 

Setting the reflectance, $\eta=\frac{1}{3}$, the input state of the meter is required to be:
\begin{equation}
|D'\rangle\equiv\big|D(1/3)\rangle=\frac{\sqrt{3}}{2}|H\rangle_m+\frac{1}{2}|V\rangle_m,
\end{equation}
and the probability of success for the two eigenstate signal inputs is $P_V=\frac{1}{2}$ and $P_H=\frac{1}{6}$, respectively. For an arbitrary input $\alpha|H\rangle_s+\beta|V\rangle_s$, the probability of success is a weighted average of the two:
\begin{equation}
P_{\alpha|H\rangle_s+\beta|V\rangle_s}=\frac{\alpha^2+3\beta^2}{6}.
\end{equation}
Success of the circuit is heralded by the detection of a single photon in the meter output, i.e., one photon at one of the two detectors.
The success probability is made equal for all input states by introducing the $\frac{2}{3}$ loss (a $\frac{2}{3}$BS), as indicated in Fig.~\ref {schematic}, however an additional detector is then required to monitor this extra output port to check that no photon exits there. 

In summary: for a vertical signal state, $|V\rangle_s$, the unequal superposition state of the meter (Eq.~\ref{d}) combined with the $\frac{2}{3}$ loss experienced by the $H$ component produces $|D\rangle_m$ which is then rotated by a waveplate to $|V\rangle_m$. For a horizontal signal state, $|H\rangle_s$, the two photon non-classical interference at the $\frac{1}{3}$BS combined with single photon detection of the meter, induces a $\pi$ phase shift in the $H$ component. This phase shift is transfered to the polarization state of the meter to produce the output $|A\rangle_m$, which is then rotated by the waveplate to $|H\rangle_m$. Finally, note that for the signal in the state $|H\rangle_s$ we make a QND measurement of the photon's presence or absence by measuring the polarization of the meter. 

Although closely related to the CNOT gate of Ref.\cite{job03} there are a couple of distinct features of the QND gate which we wish to highlight: 
\begin{enumerate}
\item Because the target (meter) is in a known state the loss present in the target arm for the full non-deterministic CNOT can be avoided thus enhancing the success probability of the gate from 1/9 to 1/6 (when signal loss is included);
\item The operation of the full CNOT gate is post-selected. However, it is in principle possible to obtain heralded operation from the QND gate by requiring one and only one photon be detected at the meter output (and no photons at the signal loss port, if present).
\end{enumerate}

Because the optical QND gate is based on a CNOT gate generalized measurements can also be implemented. This is correct. As well as performing QND measurements of photon polarisation, the device shown in Fig. \ref{schematic} can also be used to make non-destructive, arbitrary strength measurements of polarisation. This is achieved by varying the input state of the meter. Consider what happens when the meter input is $|V\rangle_m$: regardless of the signal polarisation, the meter photon travels through the $m_V$ mode and no measurement of the signal is made. A measurement of the signal photon of any strength between no measurement and a projective measurement can be made by preparing the meter in the appropriate real superposition 
\begin{equation}
|\psi\rangle_m=a|H\rangle_m+\sqrt{1-a^2}|V\rangle_m; 
\label{m_weak}
\end{equation}
$a\in[0,\sqrt{3}/2]$ (ie. $|\psi\rangle_m\in\{|D'\rangle_m\rightarrow|V\rangle_m\}$).
In this arbitrary strength measurement regime, it is necessary to introduce the additional $\frac{2}{3}$BS into the signal arm to balance the amplitudes in $s_{Vo}$ and $s_{Ho}$. Because the measurement is no longer projective, unequal amplitudes in these modes would cause the signal state to be rotated, rather than simply cause the success probabilities to be input state dependent.

\subsection{Characterising non-deterministic, QND coincidence measurements}

Our discussion of measures for characterising QND measurements in section \ref{section2} implicitly assumed that the measurements were deterministic. However the optical implementation we have been discussing in this section is non-deterministic. How should it be characterised? It seems reasonable that the characterisation should be based on only those events for which the device has been predicted to work: when a single meter photon is found after a single signal and meter photon are injected into the device. This question becomes a bit more subtle for the actual experimental situation \cite{Pry04} because these events can not be unambiguously identified until both the meter {\it and} the signal photons have been detected - coincidence detection. Thus the characterisation in \cite{Pry04} was based only on the post-selected events in which photons arrived at both the meter and signal outputs. It has been suggested in Ref.~\cite{kok04} that this post-selection is not justified. Here we argue that it is a sensible application of the figures of merit.
	
Firstly; the claim of this section is that if two photons are simultaneously incident at the meter and signal inputs of the gate and one and only one photon arrives at the meter output (and no photon departs through the signal loss port if present) then the gate performs a QND measurement of the signal's polarization. However, experimentally it is not possible to reliably post-select successful events based only on the meter output. This is because of technical limitations associated with source and detector efficiency (it is not sure that a photon was present in both the signal and meter inputs) as well as the threshold nature of the detectors (they can only discriminate between zero and more than one photons). By looking at both meter and signal outputs it is possible to reliably post select only those events corresponding to the theoretical description, as was done in Ref.~\cite{Pry04}, where the figures of merit confirmed the in principle operation of the gate. Currently, virtually all photonic quantum information demonstrations are of this type. The description of the signal as free-propagating after the QND measurement is valid in the sense that any transformation can be performed on the signal output (including further QND measurements), provided, in the end, only those events where a photon is eventually counted are post-selected.

Secondly; from an operational point of view, in spite of the limitations of the current experiments, the system can still achieve the goals of a more sophisticated QND device in particular situations. Consider a Quantum Key Distribution scheme in which an eavesdropper, Eve, uses the system to make a QND attack on the line. Eve is only ever interested in her meter results in those situations in which (i) she detected a photon and (ii) Bob detected a photon. She knows which events these are from her own records and from analyzing Alice and Bob's public discussion after key exchange. These are precisely the events we suggest are used to calculate the fidelities and so these are the correct numbers to characterize the effectiveness of her attack in this situation.

We conclude that characterisation of a non-deterministic QND gate through coincidence measurements has a clear in-principle and operational interpretation.

\section{Measurements of Weak Values} 
\label{weak}
\subsection{Weak Measurements and Weak Values} 

As discussed in section \ref{cnot}, a CNOT gate allows a QND measurement to be performed with a variable strength, as measured by the 
measurement fidelity $F_{\rm M} =\gamma^2$ for eigenstate inputs.  As before 
$\gamma$ lies between $1/\sqrt{2}$ and $1$ and parametrizes the probe input state $\gamma 
\ket{0} + \bar{\gamma}\ket{1}$, where $\bar{\gamma} \equiv 
\sqrt{1-\gamma^2}$.  The limit $\gamma = 1$ correspond to a strong, or 
projective measurement of the QND observable.  Such a measurement collapses the system 
into an eigenstate.  With $1/\sqrt{2} < \gamma < 1$ the measurement is not 
projective, and the disturbance to the system is reduced.  The limit $\gamma 
\to 1/\sqrt{2}$ is that of {\em weak measurements} where the amount of 
information obtained is arbitrarily small, and the disturbance is also 
arbitrarily small.

Weak measurements are of fundamental interest because 
they allow one to measure a {\em weak value}.  A weak value is just the 
mean value of a weak measurement.  That is, it is obtained by averaging 
over a large ensemble of weak measurement results on identically prepared 
systems, just as is the mean value of a strong measurement.  However, 
because of the imprecision in each weak measurement result, the size of the 
ensemble must be correspondingly larger than in the case of strong 
measurements in order to obtain reliable results.  

Simply considering a prepared state $\ket{\psi}$ gives an 
uninteresting weak value --- the same as the strong value for the same 
quantity: 
\beq 
\an{X_{\rm weak}}_{\psi} = \an{X_{\rm strong}}_{\psi} = 
\bra{\psi}\hat{X}\ket{\psi}.  
\eeq
As realized by Aharonov, Albert, and 
Vaidman \cite{AhaAlbVai88}, {\em post-selection} can lead to interesting and non-intuitive weak values that differ from the corresponding strong value.  That is, the average is calculated from the 
sub-ensemble where a {\em later} strong measurement reveals the state to be 
$\ket{\phi}$.  The post-selected weak value is found to be 
\beq 
\label{weakvalgen} 
\bbb{\phi}\an{X_{\rm weak}}_{\psi} = {\rm 
Re}\frac{\bra{\phi}\hat{X}\ket{\psi}}{\ip{\phi}{\psi}}.  
\eeq

This 
expression is unusual, in that the numerator and denominator are linear in 
$\ket{\psi}$ and $\ket{\phi}$ rather than bilinear.  This has the 
consequence that the weak value can lie {\em outside} the range of 
eigenvalues of $\hat{X}$ \cite{AhaAlbVai88}.  This was soon verified 
experimentally \cite{RitStoHul91}.  However, it is worth remarking that in \cite{RitStoHul91} and 
later experiments the measurement device used to probe the system was 
actually another degree of freedom of the system.  Thus the experiment can 
be interpreted within single-particle quantum mechanics (i.e. it displays only semi-classical statistics). In contrast, a weak 
value experiment performed using the variable strength QND measurements we have described in section \ref{cnot} requires 2 particle entanglement.  
Weak values have now been used to analyse a great variety of quantum 
phenomena \cite{Ste95,Aha02,Mol01,Wis02a,RohAha02,Brun03,Sol04,Gar04}.  

A mean 
value of an observable $\hat{X}$ outside its eigenvalue range 
cannot occur for a strong measurement of $\hat{X}$, even if the results 
are post-selected.  This constraint can be seen explicitly from the expression for the 
post-selected strong value, 
\beq 
\label{strongvalgen} 
\bbb{\phi}\an{X_{\rm strong}}_{\psi} = 
\frac{\sum_{x} \st{\ip{\phi}{x}}^{2} x\st{\ip{x}{\psi}}^{2}} {\sum_{x'} 
\st{\ip{\phi}{x'}}^{2} \st{\ip{x'}{\psi}}^{2}}.  
\eeq
The denominator in this expression 
 is the probability $P(\phi|\psi)$ to obtain the final 
measurement result $\phi$ irrespective of the result of the intermediate 
measurement of $\hat{X}$.  

From this expression it is also obvious that the 
average result of the strong measurement of $\hat{X}$, summed over all 
final measurement results is the non-post-selected result, 
\bqa \sum_\phi 
\bbb{\phi}\an{X_{\rm strong}}_{\psi} P(\phi|\psi) &=& \sum_\phi \sum_{x} 
\st{\ip{\phi}{x}}^{2} x\st{\ip{x}{\psi}}^{2}\nn \\ &=& \an{X}_\psi, 
\eqa 
where $\cu{\ket{\phi}}$ is a complete set of orthonormal states.  It is 
less obvious, but also true, that this result holds for weak measurements.  
In that case, the initial state is hardly disturbed by the measurement of 
$\hat{X}$ so that $P(\phi|\psi) = \st{\ip{\phi}{\psi}}^2$ and 
\bqa 
\sum_\phi \bbb{\phi}\an{X_{\rm weak}}_{\psi} P(\phi|\psi) &=& \sum_\phi 
{\rm Re}\ro{\ip{\psi}{\phi}\bra{\phi}\hat{X}\ket{\psi}} \nn\\ 
&=& 
\an{X}_\psi.  
\eqa 

\subsection{Weak Values for a Qubit} 

Consider now measuring the logical state of a qubit using a QND device.  Because 
physically this is realized as the photon number (zero or one) of some 
mode, we will call this observable $\hat{n}$.  For simplicity let us consider a single-rail qubit, prepared in 
the state $\ket{\psi} = \alpha\ket{0}+ \beta\ket{1}$, where the value of zero or one in the ket represents both its occupation number and its logical value \cite{lund} (as distinct from the dual-rail logic described in section \ref{optics}).
Say we were to post-select the weak measurement results on a final (strong) 
measurement of the qubit in the logical basis, yielding result $m \in 
\cu{0,1}$.  Then \erf{weakvalgen} and \erf{strongvalgen}  show that the weak value would again be 
the same as the strong value, and would agree with the final result: 
\beq 
\bbb{m}\an{n_{\rm weak}}_\psi = \bbb{m}\an{n_{\rm strong}} _\psi = m.  
\eeq

To obtain an interesting weak value, outside the eigenvalue range of 
$\{0,1\}$, it is necessary to make the final measurement in a basis 
different from that of the weak measurement.  This motivates considering a 
final measurement in a basis conjugate to the logical basis.  Without loss 
of generality we can then consider the basis $\ket{\pm} = (\ket{0}\pm 
\ket{1})/\sqrt{2}$, and the weak value conditioned on the result $+$, 
\beq 
\label{plusweakpsi} 
\bbb{+}\an{n_{\rm weak}}_\psi = {\rm 
Re}\frac{\beta}{\alpha+\beta}.  
\eeq 
Since we can choose $\alpha \approx 
-\beta$, Eq.~\ref{plusweakpsi}  can take any value on the real line.  

In an 
experiment, the weak measurement cannot be arbitrarily weak. Thus there will 
be corrections to \erf{plusweakpsi} due to the finiteness of $\gamma 
-1/\sqrt{2}$.  Consider first the case of non-post-selected measurements.  
As shown in Sec.~\ref{cnot}, the entangled state of the system and meter is 
\beq \label{ent} 
(\alpha\gamma\ket{0}_s+\beta\bar\gamma\ket{1}_s )\ket{0}_m 
+ (\alpha\bar\gamma\ket{0}_s+\beta\gamma\ket{1}_s )\ket{1}_m 
\eeq 
Thus the 
probability of measuring the meter to be in state $k=0$ or $1$ is 
\beq 
P(k|\psi) = \bra{\psi} \hat{E}_k \ket{\psi},
\eeq 
where 
\beq 
2\hat{E}_k = 
\hat{1} - (-1)^k (2\gamma^2-1) (2\hat{n} - \hat 1 ) 
\eeq 
Thus the mean 
value of $\hat{n}$ can be determined from the measurement results via 
\beq 
2\an{n}_\psi - 1 = \frac{P(1|\psi) - P(0|\psi)}{2\gamma^2 - 1} .  
\eeq 

Now 
with post-selection, the mean photon number given by the weak measurement will be 
\beq 
2\bbb{\phi}\an{n}_\psi - 1 = \frac{P(1|\phi,\psi) - 
P(0|\phi,\psi)}{2\gamma^2 - 1} .  
\eeq 
From \erf{ent}, these probabilities 
are 
\bqa P(0|+,\psi) &=&{{1}\over{2}} |\alpha\gamma + \beta\bar\gamma|^2/ P(+|\psi), \\ 
P(1|+,\psi) &=& {{1}\over{2}}|\alpha\bar\gamma + \beta\gamma|^2 / P(+|\psi), 
\eqa 
where 
$P(+|\psi) =( |\alpha\gamma + \beta\bar\gamma|^2 + |\alpha\bar\gamma + 
\beta\gamma|^2)/2 = (1 + 4\gamma\bar{\gamma}{\rm Re}[\alpha\beta])/2$.  This gives 
\beq \label{plusarbpsi} 
\bbb{+}\an{n}_\psi = \frac{|\beta|^2 + 
2\gamma\bar\gamma{\rm Re}[\alpha\beta^*]}{1+4\gamma\bar\gamma{\rm 
Re}[\alpha \beta]}.  
\eeq 
It can be shown that the corresponding result for 
post-selecting on finding the system in state $\ket{-}$ is such that 
\beq 
P(+|\psi) \bbb{+}\an{n}_\psi + P(-|\psi) \bbb{-}\an{n}_\psi = |\beta|^2.  
\eeq 
That is, the formalism makes sense for arbitrary strength measurements 
(arbitrary $\gamma$), not just strong and weak measurements.  

In the strong 
measurement limit, $\gamma \to 1$, so that $\bar\gamma \to 0$, this 
evaluates to 
\beq 
\bbb{+}\an{n_{\rm strong}}_\psi = |\beta|^2,  
\eeq 
which 
is as expected because the final measurement is conjugate to the strong 
measurement of $\hat{n}$, so there are no correlations between the final 
measurement result and the QND measurement result and the post-selection 
has no effect.  

In the weak measurement limit with $\gamma \to 1/\sqrt{2}$, 
then as long as $\alpha -\beta \neq 0$, \erf{plusarbpsi} evaluates to 
\erf{plusweakpsi}.  In a real experiment $\gamma$ must be finitely greater 
than $1/\sqrt{2}$, otherwise it would take an infinite ensemble size to 
obtain sufficient data to produce a reliable mean value for the weak 
measurement.  It is therefore of interest to know how strong the 
measurement can be (how large $\gamma$ can be) while still yielding an 
interesting weak value (i.e.  a negative weak-valued mean photon number).  
Assuming for simplicity that $\alpha$ and $\beta$ are real, with $ 
1/\sqrt{2} < \alpha < 1$ and $ -1/\sqrt{2} < \beta < 0$, it is easy to show 
that \erf{plusarbpsi} is negative as long as 
\beq 
\frac{1}{2} < \gamma^2 < 
\frac{1}{2}\ro{1+ \sqrt{2\alpha^2 - 1}/\alpha }.  
\eeq 
The closer $\alpha$ 
is to $1/\sqrt{2}$, the more stringent becomes the weakness requirement on 
$\gamma$.  For example, if $\alpha = 0.8$ and $\beta = -0.6$, then the 
above inequality gives $\gamma < 0.911$.  If we choose $\gamma = 0.8$ then 
\erf{plusarbpsi} gives $\bbb{+}\an{n}_\psi = -9/7$.

\section{Conclusion}

In this paper we have explored QND measurements, focussing particularly on qubits and using examples from optics. We introduced general figures of merit for QND operation based on classical fidelities between the measured distributions of the various inputs and outputs of the device. These can be applied regardless of the Hilbert space dimensions. We discussed bounds on the back action a QND device produces on the conjugate to the QND observable and defined a coherent QND measurement as one that saturates those bounds. As an abstract example we considered qubit QND measurements carried out using a CNOT gate. We showed that this was an example of an ideal QND device, both for projective measurements and generalized measurements of arbitrary strength. As a physical example we considered a non-deterministic optical realization and discussed its characterization under realistic conditions.  Finally we looked at a fundamental application of qubit QND: the measurement of weak values. We predict that weak expectation values lying well outside the eigenvalue range of the qubit observable could be obtained using a CNOT gate.

This research was supported by the Australian Research Council.

\end{document}